# Prediction on total inventory of radioisotopes produced by the interaction of $^{20}$Ne beams on $^{181}$Ta (110-170 MeV)

*Sumana* Mukherjee[1], *Susanta* Lahiri [1,2,3]*, *Rajarshi* Raut[4], *and Chiranjib* Barman[1+]

[1]Department of Physics, Sidho-Kanho-Birsha University, Purulia-723104, India
[2]Department of Chemistry, Diamond Harbour Women's University, South 24 Pgs-743368, India
[3]Department of Chemistry, Rani Rashmoni Green University, Hooghly-712409, India
[4]UGC–DAE CSR Kolkata centre, Bidhan Nagar, Kolkata, 700106, India

*Corresponding author: susanta.lahiri.sinp@gmail.com,
[+]Co-corresponding author: chiranjib-barman@skbu.ac.in

**Abstract.** This study aims to model the evaporation residues produced by the interaction of 108-170 MeV of $^{20}$Ne with the $^{181}$Ta target using Monte-Carlo PACE4 simulation code. Reaction mechanisms taking place in these interactions along with production cross sections have been discussed. Possibility of the production of clinically important exotic neutron deficient radioisotopes from these interactions have been assessed, but the result is not encouraging for production of radioisotopes useful in nuclear medicine.

## 1 INTRODUCTION

The dimension of nuclear physics is changing rapidly with the advancements in predictive modelling, advanced computational simulations and close integration of theory and experiment. In the global context, nuclear scientists are working to reach the unreachable corners of the radionuclide chart. Use of the high energy intense flux of heavy ion (HI) beams and secondary ion beams are preferred ways to reach upper part of the radionuclide chart. In the past few decades converter targets (CT) are coming up as competent and versatile target materials from different aspects. Worldwide efforts are being build up for mega facilities with CT to meet the demand for exotic radioisotopes. These targets are high Z target materials having certain physical and chemical properties like ease of heat transfer, high melting point, etc. The example of converter targets are Liquid Mercury, Lead Bismuth Eutectic (LBE), Lead Lithium Eutectic (LLE), Lead Magnesium Eutectic (LME), and the refractory metals, Tantalum, Tungsten, etc. Converter targets have also been proposed as target material in accelerator driven systems (ADS), neutron spallation source, etc. [1,2].

In CERN-ISOLDE users meeting of 2014 Lahiri et.al. [3] proposed that in addition to serving as a basic source of fundamental scientific research, the converter targets may also serve as a significant and robust source of radioisotopes, many of which could be used as novel Positron Emission Tomography (PET) radioisotopes, or therapeutic/theragnostic radioisotopes. In this direction, Maiti et al. [4] and Choudhury et al. [5] made total inventory of radioisotopes produced by the interaction of 1.4 GeV proton induced reactions with LBE target.

There are rare reports on the production of radioisotopes by the interaction of HI beams with converter targets. The Variable Energy Cyclotron Centre (VECC), Kolkata, India is now equipped with $^{20}$Ne beam. Therefore, in the present work an attempt has been made to predict the total inventory of radioisotopes produced by the interaction of $^{20}$Ne beams with tantalum targets. The rationale for choosing tantalum as target material is its use as converter target and also as container when liquid or molten targets are used as converter material. The reports are available in literature depicting the interactions of different projectiles at various energies with tantalum target for studying nuclear dynamics, reaction products, fission, etc. However, to the best of our knowledge, there is no report on the total inventory of radioisotopes produced from the interaction of tantalum target with $^{20}$Ne at the projectile energy range reported in this paper. The prediction has been carried out by Monte Caro simulation code PACE (Projection Angular-momentum Coupled Evaporation).

The purpose is to predict the isotopes resulting from the reaction of tantalum with neon at 110-170 MeV and also search for some clinically important radioisotope.

## 2 Methodology

### 2.1 PACE

The PACE4 code has a wide range of applications, including the calculation of fission barriers, reaction cross sections, recoil velocities, etc., and is particularly well suited for HI induced nuclear reactions [7]. PACE4 is based on equilibrium (EQ) reaction mechanism. In this approach, consecutive Monte Carlo simulations are used to tackle the de-excitation of the compound nucleus. The Bass formula is used to compute the cross sections of evaporation residues [8]. Proton, neutron, and α-particle emission optical model characteristics are taken from the report by Perey and Perey [9].

The diffuseness of the partial-wave distribution (INPUT) has been adjusted to 1 in the current computations. FYRST, the level density option, was set to 0. Additionally, the modified rotating liquid-drop fission barrier provided by A. J. Sierk [10] has been incorporated into the parameter BARFAC, which was set to 0. The parameter FACLA has been taken as 10, and the ratio of the Fermi-gas level density parameter at the saddle point to that at the ground state has been assumed to be unity. During the de-excitation process, the algorithm internally calculates the nuclear masses and level densities. With an adjustable fission barrier, fission is considered as a decay mode.

Total 20000 number of cascades have been chosen for each computation. Natural Ta ($^{181}$Ta = 99.98799% abundance [11]) has been taken as target material. The beam energies were chosen as 110, 125, 140, 156, 170 MeV, (the upper limit for $^{20}$Ne beam at VECC, Kolkata. The target thickness has been considered as 4 mg cm$^{-2}$, the exit energies were calculated using SRIM (stopping and range of ion in matter) [11].

## 3 Result and Discussions

The PACE4 calculation for $^{181}$Ta+$^{20}$Ne reactions at 110-170 MeV projectile energies predicts total 50 numbers of isotopes with mass number varying from 183-198. Amongst these 37 isotopes have cross section > 1mb and 31 isotopes have half-life > 1 min. The nuclear characteristics of the radioisotopes having production cross section >1mb and $T_{1/2}$> 1 min have been listed in Tabe 1. Many of these isotopes, e.g., $^{186,187}$Pt, $^{188}$Au, $^{189,190}$Hg, $^{189,190,191,192,193}$Tl, $^{191,192}$Pb,

[194]Bi have no well-defined decay data as per NUDAT3 [11]. The production cross-sections for the rest of the isotopes in all five projectile energies have been provided in Table 2.

**Table 1.** Radioisotopes produced by $^{181}$Ta+$^{20}$Ne interactions
(only the radioisotopes having production cross section >1mb and $T_{1/2}$> 1 min have been listed)

| Radio-isotope | Half- life | Maximum cross section (mb) @ projectile energy (MeV) | Decay mode | Principal $\gamma$ energy, keV (intensity, %) | Particle energy ($\beta^{-}/\alpha$), keV | Auger electron, keV (intensity, %) |
|---|---|---|---|---|---|---|
| $^{184}$Ir | 3 h | 1.0@170 | ε (100%) | 119.7 (30.8)<br>263.8 (64.8)<br>511(24.7) | | 6.8 (91.9)<br>48.3 (4.2) |
| $^{186}$Pt | 2 h | 3.0@170 | ε+β+ (99.9%) | Decay data not Available | | |
| $^{187}$Pt | 2.3 h | 2.2@170 | ε+β+ (100%) | Decay data not Available | | |
| $^{186}$Au | 10.7 min | 13.7@170 | ε (100%) | 191.5 (62.4)<br>298.8 (25.6)<br>511.0 (102) | | 7.2 (42.7) |
| $^{187}$Au | 8.3 min | 12.@170 | ε+β+ (100%) | 511.0 (13.8)<br>1266.5 (3.1) | | 7.2 (70.4) |
| $^{188}$Au | 8.8 min | 5.7@156 | ε+β+ (100%) | Decay data not Available | | |
| $^{189}$Au | 4.6 min | 2.1@140 | ε (100%) | 166.4 (59.0)<br>511.0 (20.0) | | |
| $^{190}$Au | 43.0 min | 2.3@170 | ε (100%) | 295.8 (90.0)<br>511.0 (16.00 | | 7.2 (55.6) |
| $^{188}$Hg | 3.0 min | 8.5@170 | ε (99.9%) | 190.1 (3.6)<br>511.0 (2.7) | | 7.4 (104) |
| $^{189}$Hg | 8.0 min | 52.3@170 | ε+β+ (100%) | Decay data not Available | | |
| $^{190}$Hg | 20.0 min | 49.1@156 | ε+β+ (100%) | Decay data not Available | | |
| $^{191}$Hg | 50.8 min | 15.8@140 | ε+β+ (100%) | 252.6 (55.0)<br>511.0 (7.4) | | 7.4 (111.0) |
| $^{192}$Hg | 4.8 h | 8.4@140 | ε (100%) | 157.2 (7.2)<br>274.8 (52.0) | | 7.4 (105)<br>52.4 (4.5) |
| $^{189}$Tl | 2.0 min | 145.0@170 | ε+β+ (100%) | Decay data not Available | | |
| $^{190}$Tl | 2.8 min | 134.0@156 | ε+β+ (100%) | Decay data not Available | | |
| $^{191}$Tl | 5.0 min | 135.0@140 | ε+β+ (100%) | Decay data not Available | | |
| $^{192}$Tl | 10.0 min | 80.2@125 | ε+β+ (100%) | Decay data not Available | | |
| $^{193}$Tl | 22.0 min | 35.2@156 | ε+β+ (100%) | Decay data not Available | | |
| $^{196}$Tl | 1.8 h | 1.6@125 | ε (100%) | 425.7 (83.0)<br>511.0 (25.0)<br>1495.8 (8.1) | | 7.6 (48) |
| $^{191}$Pb | 1.3 min | 23.6@170 | ε+β+ (99.9%) | Decay data not Available | | |
| $^{192}$Pb | 3.5 min | 294.0@170 | ε+β+ (99.9%) | Decay data not Available | | |
| $^{193}$Pb | 5.8 min | 215.0@156 | ε+β+ (100%) | 392.2 (68.3)<br>511.0 (55.3)<br>716.5 (22.1) | | 7.7 (46.1) |
| $^{194}$Pb | 10.7 min | 91.7@156 | ε (100%) | 203.8 (16.6)<br>581.8 (19.3)<br>1519.4 (16.8) | | 7.7 (70.6) |
| $^{195}$Pb | 15.0 min | 148@125 | ε (100%) | 383.6 (91.0)<br>394.2 (44.2)<br>511.0 (23.1) | | 7.7 (101.0) |

Table 1 (contd.)

| Radio-isotope | Half- life | Maximum cross section (mb) @ projectile energy | Decay mode | Principal $\gamma$ energy, keV (intensity, %) | Particle energy ($\beta^-$ /$\alpha$), keV | Auger electron, keV (intensity, %) |
|---|---|---|---|---|---|---|
| $^{196}$Pb | 37.0 min | 80.0@ 110 | ε (100%) | 253.1 (27.0)<br>502.1 (27.0) | | 7.7 (76.0) |
| $^{197}$Pb | 8.0 min | 6.9@110 | ε (100%) | 385.8 (50.0)<br>511.0 (7.9)<br>761.1 (13.0) | | 7.7 (53.0) |
| $^{193}$Bi | 1.0 min | 148@156 | ε+β+ (96.5%)<br>α (3.5%) | | $E_\alpha$,<br>5899<br>(2.1) | |
| $^{194}$Bi | 1.6 min | 210@140 | ε+β+<br>(99.54%)<br>α (0.46%) | Decay data not Available | | |
| $^{195}$Bi | 1.4 min | 281.0@125 | α (100%) | - | $E_\alpha$,<br>6106<br>(33) | |
| $^{196}$Bi | 5.0 min | 273.0@110 | ε (100%) | 511.0 (107.0)<br>689.3 (35.5)<br>1049.4 (87.0) | | 7.9 (27.1) |
| $^{197}$Bi | 5.0 min | 17.1@110 | α (100%) | - | $E_\alpha$,<br>5776<br>(60.0) | |

**Table 2:** Cross sections of different radioisotopes at different projectile energies

| Radioisotope | Half-life | Cross section (mb) | | | | |
|---|---|---|---|---|---|---|
| | | 110 MeV | 125 MeV | 140 MeV | 156 MeV | 170 MeV |
| $^{184}$Ir | 3 h | - | - | - | 0.4 | 1.0 |
| $^{186}$Au | 10.7 min | - | - | - | 0.4 | 13.7 |
| $^{187}$Au | 8.3 min | - | - | 0.3 | 8.8 | 12.1 |
| $^{189}$Au | 4.6 min | - | 0.9 | 2.1 | 0.8 | 2.0 |
| $^{190}$Au | 43.0 min | 0.2 | 0.5 | 0.2 | 0.8 | 2.3 |
| $^{188}$Hg | 3.0 min | - | - | - | - | 8.5 |
| $^{191}$Hg | 50.8 min | - | 0.3 | 15.8 | 13.7 | 2.0 |
| $^{192}$Hg | 4.8 h | 0.0 | 6.4 | 8.4 | 0.7 | 2.1 |
| $^{196}$Tl | 1.8 h | 1.1 | 1.6 | 0.1 | - | - |
| $^{193}$Pb | 5.8 min | - | - | 26.6 | 215 | 75.9 |
| $^{194}$Pb | 10.7 min | - | 33.9 | 278 | 91.7 | 7.5 |
| $^{195}$Pb | 15.0 min | 5.9 | 148 | 34.5 | 1.3 | - |
| $^{196}$Pb | 37.0 min | 80.0 | 36.6 | 1.1 | - | - |
| $^{197}$Pb | 8.0 min | 6.9 | 0.1 | 1.1 | - | - |
| $^{193}$Bi | 1.6 min | | | 28.5 | 148.0 | 35.8 |
| $^{195}$Bi | 1.4 min | | 1.06 | 210.0 | 48.0 | 2.7 |
| $^{196}$Bi | 5.0 min | 273.0 | 59.3 | 0.7 | - | - |
| $^{197}$Bi | 5.0 min | 17.1 | 2.3 | | | |

It is interesting to note that as PACE4 follows the EQ reaction mechanism, therefore it predicts the heavier mass products at lower energies. At higher bombarding energies more fragments take place. The excitation functions of different radioisotopes of lead has been plotted in Figure 1. The same for Hg and Bi isotopes has been potted in figure 2.

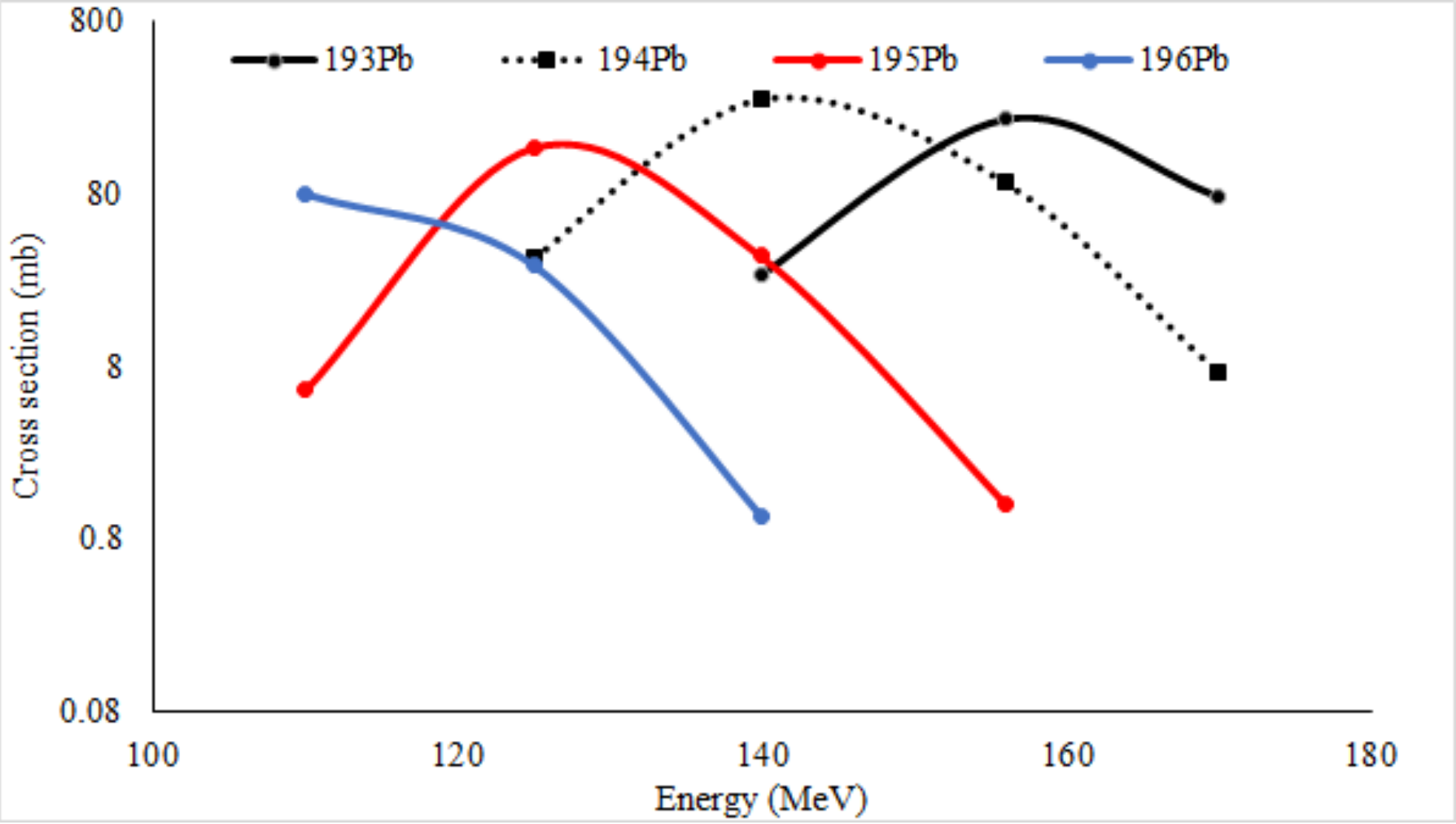

**Fig. 1.** Excitation functions of different Pb isotopes.

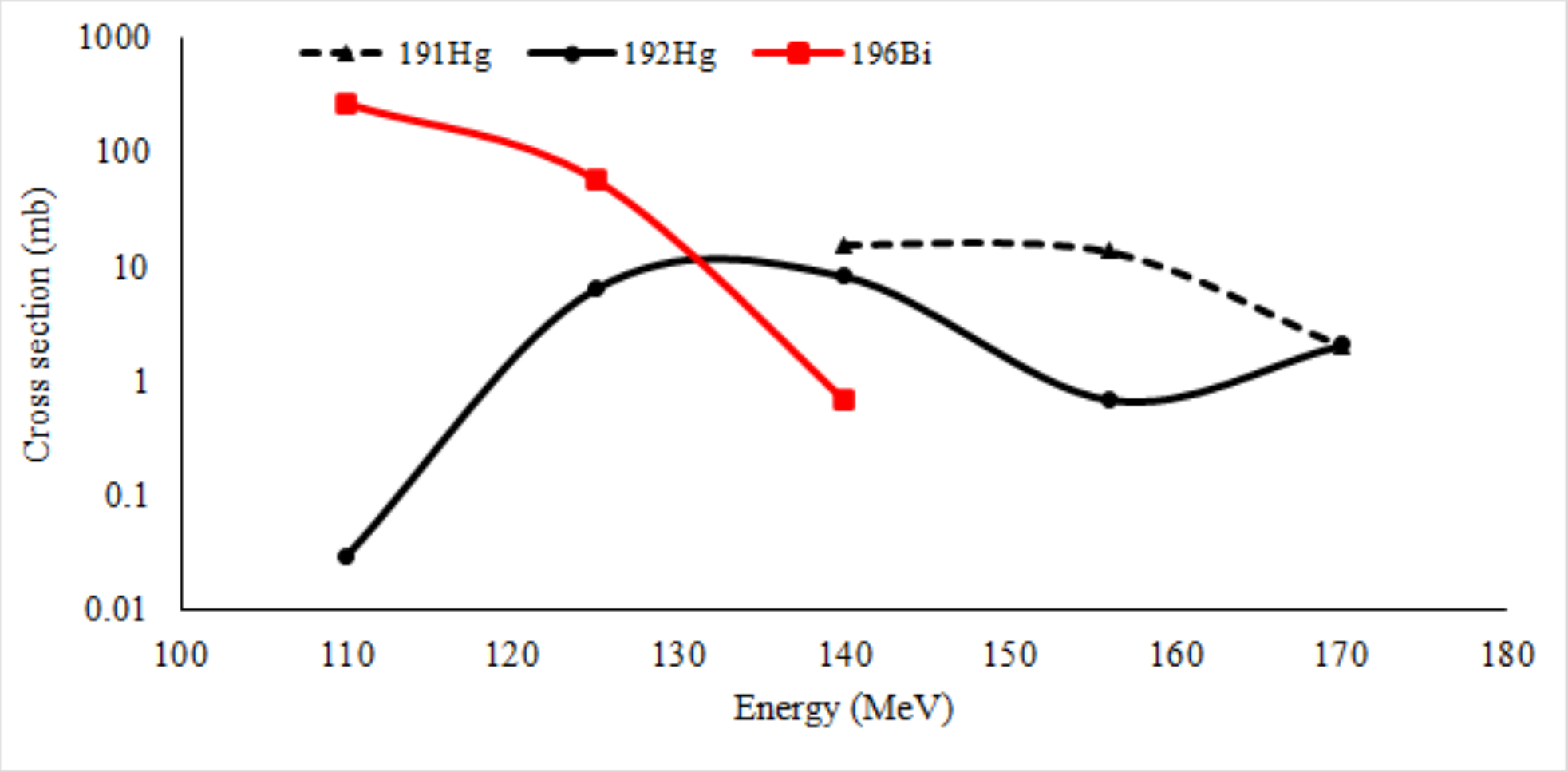

**Fig. 2**. Excitation function of different Hg and Bi radioisotopes.

Careful look on Table 1 and Table 2 shows that there are some radionuclides qualify as clinically important. For example, $^{184}$Ir, $^{196}$Tl, have half-lives in hour scale and possess good intensity of 511.0 keV $\gamma$. Therefore these radionuclides may be useful in PET diagnosys. However, the production cross-sections of these radionuclides are too low for human application. On the other hand, $^{193\text{-}196}$Pb radionuclides have comparatively better cross-sections, but their half-lives are too low to deal with human application. $^{195}$Bi and $^{197}$Bi are high energy $\alpha$ emitters, but their half-lives are too short for therapeutic application.

## 4 Concusions

This paper enlists the evaporation residues raised from the $^{20}$Ne + $^{181}$Ta nuclear reaction in the projectile energy range 110-170 MeV. The Monte Carlo simulation code PACE4 considers only EQ reactions. Therefore if the DIR reactions occur in this projectile energy range, that has not been predicted by PACE4. The actual validation both qualitative and quantitative will come from performing the experiment in the same projectile energy range. Nevertheless, this work would give a fair estimation of the evaporation residues.